# Large-scale intermittency and rare events boosted at dimensional crossover in anisotropic turbulence

Keiko Takahasi[1], Koji Goto[2], Ryo Onishi[1], Masatoshi Imada[3]
**Affiliations:**
[1] Center for Earth Information Science and Technology, Japan Agency for Marine-earth Science and Technology, 3173-25 Showa-machi, Kanazawa-ku, Yokohama 236-0001, Japan
[2] 1st Government and Public Solutions Division, NEC Corporation, 7-1 Shiba 5-chome, Minato-ku, Tokyo 108-8001, Japan
[3] Department of Applied Physics, University of Tokyo, Hongo 7-3-1, Bunkyo-ku, Tokyo 113-8656, Japan

**Abstract**

**Understanding rare events in turbulence provides a basis for the science of extreme weather, for which the atmosphere is modeled by Navier-Stokes equations (NSEs). In solutions of NSEs for isotropic fluids, various quantities, such as fluid velocities, roughly follow Gaussian distributions, where extreme events are prominent only in small-scale quantities associated with the dissipation-dominating length scale or anomalous scaling regime. Using numerical simulations, this study reveals another universal promotion mechanism at much larger scales if three-dimensional fluids accompany strong two-dimensional anisotropies, as is the case in the atmosphere. The dimensional crossover between two and three dimensions generates prominent fat-tailed non-Gaussian distributions with intermittency accompanied by colossal chain-like structures with densely populated self-organized vortices (serpentinely organized vortices (SOV)). The promotion is caused by a sudden increase of the available phase space at the crossover length scale. Since the discovered intermittency can involve much larger energies than those in the conventional intermittency in small spatial scales, it governs extreme events and chaotic unpredictability in the synoptic weather system.**

## *1. Introduction*

Since fluid turbulence is chaotic, probability distributions in addition to averaged physical quantities are imperative in order to understand the turbulence deeply [1]. For the average energy flux, Kolmogorov showed that the energy injected at large (long) spatial scales cascades into smaller (shorter) scales and energy injection balances dissipation at small scales, forming an inertial range between the spatial scale of the energy injection and the dissipation range. Kolmogorov's self-similarity assumption asserts that the mean energy spectra $E(k)$, where $k$ is the wavenumber, follow universal power laws, $E(k) \propto k^{\zeta}$ with $\zeta \sim -5/3$ for three-dimensional (3D) turbulence [2] and $\zeta \sim -3$ for 2D turbulence [3] within the inertial range. These power laws have been supported by experiments and numerical simulations, at least approximately [1].

The probability density function (PDF) $P(A)$ of a macroscopic quantity $A$ is critical for understanding the chaotic and intermittent behavior of turbulence. The Gaussian distribution predicts a very low chance of extreme events. Deviations from it can be probed by the $n$-th moment $\langle A^n \rangle = \int A^n P(A) dA$, such as the flatness defined by $F(A) = \langle (A - \langle A \rangle)^4 \rangle /$ $\langle (A-\langle A\rangle)^2\rangle^2$; the intermittency associated with promoted rare-event occurrences is typically signaled by $F$ exceeding the Gaussian value 3.

Kolmogorov's claim with the assumption of self-similarity was challenged by Landau's remark [4] on the intermittent nature of the dissipation, which invalidates the self-similarity. Subsequent studies on intermittency, inspired by concepts such as anomalous scaling and multi-fractals [4-9], have proven fruitful and shown the existence of intermittency even in the inertial range. However, it should be pointed out that the intermittency (non-Gaussianity) expected in these studies emerges at small length scales, much smaller than the system size. In such small length scales or large wavenumbers $k$, the contained energy is small as well because of the Kolmogorov law $E(k) \propto k^{\zeta}$ with $\zeta$<0. Therefore, the intermittency arising from these mechanisms (anomalous scaling and dissipation) have only a limited impact on the global phenomena at least for simulations with Reynolds number available in the present computer power (in comparison to the finding under the condition of the present work as we see later including the atmospheric simulation). In fact, non-Gaussian distributions are empirically visible only for high-order spatial derivatives associated with the anomalous scaling expected for asymptotically small scales or small scales connected to the dissipation range in isotropic fluids [5-8,10-14].

The present study describes a case in which qualitatively new and strong intermittency emerges in the inertial range potentially at a much larger length scale than the scale of the dissipation and the anomalous scaling when the fluid is in a flattened 3D space, namely flat 3D fluid (F3DF), where the vertical dimension (thickness) is much smaller than the two horizontal dimensions. F3DF behaves 2D-like at scales larger than the crossover length scale (CLS), determined by the thickness, and 3D-like at smaller length scales. Unlike the large-scale intermittency associated with shear or wakes behind obstacles, the present intermittency in F3DF is not associated with large-scale forcing or boundaries. For instance, the atmosphere is a typical F3DF. Recent observations [15, 16] and numerical simulations [17, 18] have revealed the atmosphere's average statistics: the energy spectra of winds exhibit the 2D-like exponent $\zeta\sim-3$ for the scale of $O$(100~10,000 km) and the 3D-like exponent $\zeta\sim-5/3$ for scales smaller than the CLS, around $O$(100 km).

The present study demonstrates that the dimensional crossover at the CLS generates unexplored strong intermittency, further spreading to the 3D inertial range. This spatial intermittency is also corroborated by the temporal intermittency generated at the crossover time scale derived from the CLS with the aid of the spatio-temporal correspondence hypothesized by Tennekes [19]. The CLS can be much larger than the length scales of the dissipation and the asymptotic anomalous scaling regimes, thereby the intermittency induced at the CLS may involve much larger energies than the intermittency induced at those small length scales. This is indeed the case for the atmosphere, where the CLS is around 100 km and the dissipation range is below 1 km.

Our study further reveals that the CLS intermittency generates sinuous chain-like structures, which we call serpentinely organized vortices (SOV). The width and depth of the SOV structure are comparable to the CLS, while the length is much longer than the CLS as schematically illustrated in Fig.1. Inside the SOV structures, extreme events are highly promoted, which is surrounded by relatively calm areas. A SOV structure contains a mass of coherently assembled elementary vortices. Sizes of these element vortices are comparable or smaller than the CLS. The SOV structure looks like densely distributed many peas (vortices) contained in a pod (chain structure).

The finite-time Lyapunov exponent (FTLE) is a measure of chaos, in which infinitesimally small difference in the initial condition grows exponentially (see Appendix A). We find that region of large FTLE forms large-scale chain structures in the same regions with the SOV observed at the CLS intermittency, indicating that the same origin as the CLS intermittency enhances the chaos as well. Here, the origin of the intermittency is ascribed to a strong violation of the self-similarity intrinsic at the CLS.

Furthermore, a global atmospheric simulation reveals similar strong intermittency and chaos at the CLS, namely at the mesoscale $O(100\ \text{km})$, implying the universality of our finding. This finding is important because the CLS intermittency involves much larger energies than those in the asymptotically small scales responsible for the conventional intermittency in isotropic fluids and has much stronger spill-over effects, for instance those on the origin of synoptic events such as cyclogenesis.

## 2. *Direct numerical simulation of F3DF*

We performed a direct numerical simulation (DNS) for a F3DF. The continuity equation ($\nabla \cdot \boldsymbol{u} = 0$) and the following incompressible Navier-Stokes-type equation are solved for a flat cuboid with periodic boundary conditions for all three directions.

$$\frac{\partial}{\partial t}\boldsymbol{u}(\boldsymbol{x},t) = -\boldsymbol{u}(\boldsymbol{x},t)\cdot\nabla\boldsymbol{u}(\boldsymbol{x},t) - \nabla p(\boldsymbol{x},t) + \frac{1}{Re}\nabla^2\boldsymbol{u}(\boldsymbol{x},t) - \alpha_8\nabla^{-16}\boldsymbol{u}(\boldsymbol{x},t) + \boldsymbol{f}(\boldsymbol{x},t), \quad (1)$$

where $\boldsymbol{u}\left(= \left(u_x, u_y, u_z\right)\right)$ and $p$ are the velocity vector and pressure, respectively, at position $\boldsymbol{x} = (x, y, z)$ and time $t$. Reynolds number is denoted by $Re$. Note that all the variables presented in Eq. (1) are non-dimensionalized. Table 1 summarizes the computational settings. The computational domain has the spatial size of $2\pi \times 2\pi \times (2\pi/\lambda)$, where $\lambda$ is the aspect ratio. The present study mainly discusses the case with $\lambda = 64$ (i.e., N4096-λ64); two other cases (i.e., N4096-λ128 and N1024-λ32) are used to investigate the effect of the aspect ratio. The grid resolution $k_{max}\eta$, where $k_{max}$ ($= N_x/3$ in the present simulation with the two-thirds dealiasing method) is the maximum effective wavenumber and $\eta$ ($= (Re^3 \cdot \langle\varepsilon\rangle)^{-1/4}$, where $\varepsilon$ is the energy dissipation rate) is the Kolmogorov scale, is larger than 4, and is thus fine enough for investigating dissipation-scale motions. The last source term $\boldsymbol{f}(\boldsymbol{x},t)$ represents the divergence-free random force used to maintain turbulence [20, 21], through which the energy is supplied at an input rate of $\varepsilon_{\text{in}} = 4$ at input wavenumbers centered at $k_{\text{in}} = 4$ with a range of $\pm 2$ throughout this paper. The second-last term represents the 8-th order super drag with $\alpha_8 = 4$ [22]. This term is added simply to absorb the energy inversely transferred from the energy source. It should be noted that the inverse-cascade process at $k < k_{\text{in}}$ is not the main subject of this study and details of the super drag term do not alter the findings. The pseudospectral method based on the Fourier–Galerkin method was used to solve the governing equations [23]. After having confirmed that the flow reached a statistically steady state, the simulation was continued for a sufficiently long time, compared to the energy input time scale $(\varepsilon_{\text{in}}k_{\text{in}}^2)^{-1/3}$, and statistics were collected.

Figure 2 shows the energy spectra $E(k_h) \propto k_h^{\zeta}$, where $k_h\left(= \left(k_x^2 + k_y^2\right)^{\frac{1}{2}}\right)$ is the horizontal wavenumber. The spectra are the average horizontal modal spectra, computed as an average over concentric cylinders with the absolute value of $\boldsymbol{k_h}$ $\left(= (k_x, k_y)\right)$ ranging from $k_h - \frac{\Delta k}{2}$ to $k_h + \frac{\Delta k}{2}$ with any $k_z$. The step of the wavelength $\Delta k$ was set to unity. The results confirm that $\zeta \sim -\frac{5}{3}$ for large wavenumbers and $\zeta \sim -3$ for smaller wavenumbers, consistent with earlier studies [2, 21, 24]. The 2D-to-3D crossover from $\zeta = -3$ to -5/3 occurs at $k_h \sim 0.5 k_{\rm cr}$, corresponding to roughly half of the inverse of the vertical system size $L_z$, namely $k_{\rm cr} = 2\pi/L_z = k_z^{min} (= \lambda)$ [2, 24]. The observation that the crossover from $\zeta = -3$ to -5/3 occurs always at a somewhat smaller wavenumber than $k_{\rm cr}$ has been previously reported [21, 25]. A comparison between Figs. 2(a) and (b) clearly shows that the crossover from $\zeta = -3$ to -5/3 at $k \sim 0.5 k_{\rm cr}$ is a universal phenomenon, irrespective of the aspect ratio of the system.

Figure 3 shows that the horizontal velocity $u_y$ (and equivalently $u_x$) follows essentially a Gaussian distribution, as expected, whereas the horizontal velocity increment at half-height separation $\Delta u_y(L_z/2, 0) = u_y(x, y, z = L_z/2) - u_y(x, y, z = 0)$ clearly deviates, exhibiting an intermittency (note that the $x$ and $y$ dependences are omitted in the notation). We demonstrate later that this non-Gaussianity of $\Delta u_y$ emerging at the CLS increment differs from the non-Gaussianity at small increments related to the dissipation scale and the anomalous scaling regime, which is restricted to asymptotically small scale in presently available simulations [5-8]. It should also be emphasized that such non-Gaussianity of $\Delta u_y$ is not clearly visible in 3D homogeneous fluid even when one simulates by using presently available fastest supercomputers because the Reynolds number cannot be taken large enough (see such an example in Fig. B1(a)).

Figure 4 shows snapshots of the magnitude of the horizontal velocity $|\boldsymbol{u}_h| \left(= \sqrt{u_x^2 + u_y^2}\right)$ and the magnitude of the horizontal velocity increment $|\Delta \boldsymbol{u}_h| \left(= \sqrt{\Delta u_x (L_z/2, 0)^2 + \Delta u_y (L_z/2, 0)^2}\right)$. The intermittency associated with the non-Gaussianity is manifested in the snapshot for $|\Delta \boldsymbol{u}_h|$ as sparsely distributed areas that conspicuously exhibit promoted extreme events (as an assembled needle-like structure) (typically $|\Delta \boldsymbol{u}_h|/\sigma_{\Delta \boldsymbol{u}_h} > 4$) in Fig. 4(b), in contrast to the gentle structure in Fig. 4(a).

Figure 5 shows the flatness of the horizontal component of the velocity $\tilde{u}_h^K$ obtained after filtering by the spherical-shell filter (see Appendix C1), which picks up only the contribution within the wavenumber range $k - \Delta k/2 < K < k + \Delta k/2$ with a window width of $\Delta k = 1$. The flatness of $\tilde{u}_h^K$ shows a sharp peak at a filter size equal to the CLS, e.g., $K = 64$ for N4096-λ64, irrespective of the aspect ratio. The peak at $K = k_{\rm cr}$ clearly demonstrates that the present intermittency is distinct from the conventional asymptotically small-scale intermittency. Such a prominent intermittency is not visible in the inertial range in 3D homogeneous isotropic flow, as shown in Fig. B1(b).

Figure 6 shows snapshots of the $x$-$y$ plane cross section for the horizontal velocity increment, enstrophy, and FTLEs (see Appendix A), which measure unpredictability (chaos), at the same simulation time as that in Fig. 4. A remarkable feature that persisted during the simulations (see for example Figs. 6(a) and (b)) is that an intermittency consisting of prominent eddies of CLS size is assembled and forms SOV structures. Note that the raw enstrophy, i.e., even without any filter operations, exhibits the intermittency. This is due to the derivative operations applied to the velocity in the derivation of the enstrophy that give more weight to small scales, making the intermittency visible at the CLS.

Figure 6(c) indicates that in regions with such SOV structures, the Lyapunov exponent is remarkably large. Later in the discussion, we will discuss about the origin of the chaotic behavior at the SOV structure. Since the SOV structure has much longer length than the CLS, one might argue that the structure could be induced simply by the external artificial forcing. However, this is not the case. Figures 7(a-e) show the low-pass filtered horizontally oriented vorticity at the same $xy$-plane cross section as that in Fig. 6(a,b) (see Appendix C2 for the low-pass filtering). The SOV structure appears only in Figs. 7(d) and (e), where the modes with $k = k_{\rm cr}$ are included. This indicates the necessity of the CLS contribution for the SOV structure formation. It should be emphasized that the conventional asymptotically small-scale intermittency does not generate such a large coherent structure.

Figure 8 shows an example of a 3D snapshot of horizontally oriented vorticity and enstrophy. It reveals that the long SOV structures consist of a mass of assembled small spots or specks, each of which has a scale comparable to the CLS or smaller. The temporal evolution of the large SOV structures is found in Supplementary Movie S2, which shows that the large SOV structures move on a slow time scale, such as the eddy turnover time $T_e$ $(= L_x/u_{\mathrm{rms}}$, where $L_x\,(= 2\pi)$ is the horizontal size of the system and $u_{\mathrm{rms}}$ is the root mean square of the horizontal velocity).

In order to further clarify the CLS intermittency, Gaussian filtering was applied, where the weight given by the Gaussian distribution around $k = 0$ with width $1/\alpha$ is imposed as the filter (see Appendix C3). This filter can also be regarded as the real-space filter to measure the local distribution with width $\alpha$ because the Gaussian distribution is Fourier-transformed into another Gaussian in real space with the inverse width. Therefore, one can gain insight into the intermittency in real space. Figure D1 shows a peak in the Gaussian-filtered distribution at around $0.5k_{\mathrm{cr}}$. It was also confirmed that isotropic 3D fluid does not have such an intermittency except at dissipative small scales when the Reynolds number is comparable to the F3DF simulation done in the present work (see Fig. B1). All additional simulations support that the intermittency is triggered exclusively at the CLS.

The present DNS adopts periodic boundary conditions and spatial Fourier analysis, which are not applicable to usual flows in the real world. Instead, time series data at a measurement point are commonly analyzed. When the energy-containing large scale and the scale of interest are well separated, the Tennekes sweep hypothesis [19], which states that large-scale eddies advect small-scale ones, can map temporal data with time $t$ into spatial data with length scale $\ell$ via the relation $\ell = Ut$, where $U$ is the representative velocity of the system. Figure 8 shows that strong intermittency starts rising up at $k^* = \frac{2\pi}{\ell} \sim k_{\mathrm{cr}}$. This supports that the strong intermittency at the CLS is separated from the conventional small-scale intermittency [1, 5-8, 10-14]. The enhanced intermittency continues rising up above $k_{\mathrm{cr}}$ and plateaus at larger $k^*$. This spill-over effect may be the consequence of the influence of the higher-harmonic peaks in Fig. 5(a), which are broadened due to the approximate nature of the Tennekes hypothesis.

### 3. *Atmospheric simulation*

The results of a high-resolution simulation using the global atmospheric model MSSG-A [26] are described here. MSSG-A is one of the three models that participated in the global 7-km-mesh nonhydrostatic-model intercomparison project for typhoon predictions [27]. Its dynamical core is based on nonhydrostatic equations, and it predicts the three wind components, air density, and pressure. A six-category bulk cloud microphysics model is used for the equation of state for water; that is, MSSG-A is a cloud-resolving model. In contrast to the DNS for F3DF, this global simulation accounts for the effects of Earth's rotation, gravity, fluid compression/expansion, topography, moisture, and heat radiation. The five-day time integration from 00:00UTC on 13 September, 2013, performed for Typhoon Man-yi was analyzed.

In the energy spectrum in Figure 10(a), the crossover of the slope is from -3 to -5/3 at $k_h{\sim}10^{-5}$ rad/m, which corresponds to *O*(100 km), as in earlier observations [15, 16] and atmospheric simulations [17, 18]. In contrast to the equality between the CLS and the domain depth in the DNS, the atmospheric CLS is larger than the atmosphere depth *O*(10 km) because of complexities such as nonzero fluid compressibility, gravitation, Earth's rotation, and boundary conditions.

Similarly to the DNS, strong intermittency is observed at $k_h \sim 10^{-5}$ rad/m, i.e., at the CLS, in addition to the dissipation scales (Fig. 10(b)). Large structures observed in the figures of the wind increment and the Lyapunov exponents are also observed (Figs. 11(a) and (b)), in accord with the DNS of F3DF. Here, the large structures are synoptic-scale structures, e.g., tropical cyclones. Despite the various complexities of the atmospheric system, the structure of the intermittency shows remarkable similarities to the DNS results.

## 4. *Discussion*

The conclusion that the chain-like colossal structures of the assembled vortices, namely, SOV are induced at a relatively small CLS appears to break causality, because energy cascades from large to small scales. Our reasoning is as follows. A fraction of the vertically oriented vortices governing the 2D turbulence at scales larger than the CLS are transformed in the energy cascade process into horizontally oriented vortices around the CLS. There, the modes with $(k_h, k_z) = (k_{\mathrm{cr}}, 0)$ are scattered via mode coupling into same-energy modes at $k_h \sim 0$ and $k_z = k_{\mathrm{cr}}$ (Fig. 12), which necessarily generates a large horizontal structure because $k_h \sim 0$. The resultant colossal SOV structure develops a dominant rigid backbone, even at small scales, because the generated calm areas left behind at the CLS remain intact throughout the later cascade process.

We here discuss the mode coupling process in more detail. Vortices in the turbulence are generated in the cascade from large-scale structures to smaller ones through the dynamics of vortex breakup and energy flow from low to high wavenumbers. In the cascade down to the CLS, the vortices are only 2D-like because the wavenumber $k_z$ is always zero. The 3D vortex structure (horizontally oriented vortices) emerges only when a mode whose wavenumber is larger than the CLS is involved in the breakup process. Note that a horizontally oriented vortex contains fluid flow in the opposite direction between two points at the same *x,y* coordinates but different z coordinates. In the Fourier analysis, this means that a mode with nonzero $k_z$ must be involved. For instance, the vortex with a size of $L_z$ (largest 3D vortex) mainly contains the mode with $k_z = k_{\mathrm{cr}}$, which is the smallest nonzero wavenumber in the *z* direction. Therefore, it is clear that the 3D vortices are generated from the breakup of 2D vortices in the cascade beyond the CLS. During the generation of horizontally oriented 3D vortices, momentum conservation must be satisfied by the mode coupling in Navier-Stokes equations, such as the scattering process of $\boldsymbol{k}_1 = \boldsymbol{k}_2 + \boldsymbol{k}_3$ . Here, $\boldsymbol{k_1} = (\boldsymbol{k_{h1}}, k_{z1})$ satisfies $|\boldsymbol{k_{h1}}| \sim k_{\mathrm{cr}}$ and $k_{z1} = 0$ , contributing in real space to the mode of vertically oriented vortices with a radius on the scale of $1/k_{\mathrm{CLS}}$ . The scattered modes $\boldsymbol{k_2} = (\boldsymbol{k_{h2}}, k_{z2})$ and $\boldsymbol{k_3} = (\boldsymbol{k_{h3}}, k_{z3})$ must satisfy $|\boldsymbol{k_{h2}}| \sim |\boldsymbol{k_{h1}}| + p$ and $|\boldsymbol{k_{h3}}| = p$, respectively, whereas $k_{z2} = -k_{z3} = k_{\mathrm{cr}}$ (see Fig. 12). The modes $\boldsymbol{k_2}$ and $\boldsymbol{k_3}$ both contribute to horizontally oriented vortices. The process with $|\boldsymbol{k_{h3}}| \ll k_{\mathrm{cr}}$ (or equivalently $|\boldsymbol{k_{h2}}| \ll k_{\mathrm{cr}}$) contributes to the colossal structure formation with length scale $1/|\boldsymbol{k_{h3}}|$ (or $1/|\boldsymbol{k_{h2}}|$). These processes generate the horizontally oriented vortices via mode coupling in phase space expanded to the *z* direction. These mode coupling processes are clearly a consequence of the nonlinear coupling of the momenta in Navier-Stokes equations. Further discussion on mode coupling is given in Appendix E.

The interscale mode transfer from the CLS to the larger length scale, $k_h \sim k_z \sim 0$ also exists. However, the energy transfer is expected to be small. It is intriguing to clarify its role on the formation of the SOV structure in the future. Since the forcing wavenumber $k_{\mathrm{in}}$ introduces another characteristic large length scale, $k_{\mathrm{in}}$ may be involved in determining the SOV structure through the mode coupling in addition to the system size itself in the horizontal direction. The involvement of $k_{\mathrm{in}}$ for the structure and dynamics of the SOV structure is an interesting subject left for future study.

The self-similarity assumed by Kolmogorov [2] addresses that the space is active everywhere filled with vortices via the self-similar energy cascade. If the fluid is isotropic 2D, the approximate Kolmogorov law (and self-similarity) results from the nature of the vortex breakup, where after every breakup the resultant broken-up vortices fill almost the whole 2D space at each wavenumber (Fig. 13). As a result, everywhere in real space is active, and thus there is no prominent intermittency. The same is true for the case of isotropic 3D fluid.

The intermittency is generated by the breakdown of the self-similar cascade in the dissipative range and in the anomalous scaling range, where the active eddies break up into small eddies, filling only a fractal dimension smaller than the real spatial dimension [1, 9]. This generates contrasting active (extreme) and calm (inactive) regions. However, the effect of this intermittency, restricted to the dissipative range or the anomalous scaling range for isotropic fluids, is small in presently available simulations because the involved energy is limited. In contrast, when the space is suddenly expanded to the third dimension at the CLS of F3DF by including nonzero $k_z$, vortices whose size is smaller than the CLS are unable to fill the whole real space anymore because some modes escape to the 3D vortices, resulting in part of the space remaining calm without small 2D vortices (Fig. 13). Such expansion completely destroys the approximate 2D self-similarity [2, 9, 28]. The intermittent structure emerges from the contrast between the region filled with vortices (active region) and the empty region (calm region). It is important that this mechanism of the intermittency does not require the presence of dissipation and can be effective deeply inside the inertial region. The CLS scale can be much longer than the length scale of prominent anomalous scaling as well and the CLS intermittency can involve much larger energies as in the present simulation. This is a novel route to the breakdown of the self-similarity assumed by Kolmogorov. This mechanism is supported by the above-mentioned intermittency in 3D-like quantities, such as horizontal vorticity, and the vertical velocity but its absence in essentially 2D-like quantities.

Here we discuss the origin of the chaotic behavior at the SOV. Since the motion of the SOV structure is slow and continuous as is observed in Supplementary Movie S1, the global motion of the SOV may not be the origin of the chaos. We speculate that the enhanced Lyapunov exponent is caused by the internal dynamics of the SOV. Inside the SOV, the vortices are actively created and annihilated with their interactions and collisions. Dynamical process of the nonlinear excitations such as vortices and solitons is known to cause chaos [29].

The present study provides new insight into intermittency, which had mainly been explored before in connection to dissipation and anomalous scaling. A new route for strong intermittency is opened by the suddenly expanded phase space at the dimensional crossover. The mechanism of intermittency generated at the CLS and the resultant SOV together with the enhanced Lyapunov exponent may have a deep impact and connection to the synoptic structure formation and the extreme weather. This will open important future research areas about the origin and mechanism of the climate and weather dynamics as well as atmospheric phenomena, which cause disasters on the earth. Simulations within the restricted grid resolutions that do not fully cover both 2D- and 3D-like regions would fail to estimate this dominant crossover-scale intermittency. For instance, climate models that assess extreme weather are required to well resolve the present CLS intermittency. That is, they need to use at least a few 10-km or lower resolutions to resolve the $O$(100 km) CLS.

**Acknowledgments:** The authors acknowledge Yukio Kaneda and Toshiyuki Gotoh for their valuable comments. The numerical simulations were carried out on the Earth Simulator of the Japan Agency for Marine-Earth Science and Technology.

| | Aspect ratio $\lambda$ | Grids $N_x \times N_y \times N_z$ | $Re$ | $k_{max}\eta$ | TKE |
|---|---|---|---|---|---|
| N4096-λ64 | 64 | 4096×4096×64 | $1.25\times10^4$ | 4.46 | 0.506 |
| N4096-λ128 | 128 | 4096×4096×32 | $1.25\times10^4$ | 5.08 | 0.545 |
| N1024-λ32 | 32 | 1024×1024×32 | $1.00\times10^3$ | 4.34 | 1.06 |

**Table 1** Computational settings for the DNS of F3DF. $k_{max}$ (= $N_x/3$ in the present simulation) is the maximum effective wavenumber and $\eta$ is the Kolmogorov scale. TKE (=$\langle u_x^2 + u_y^2 + u_z^2\rangle/2$) is the turbulent kinetic energy.

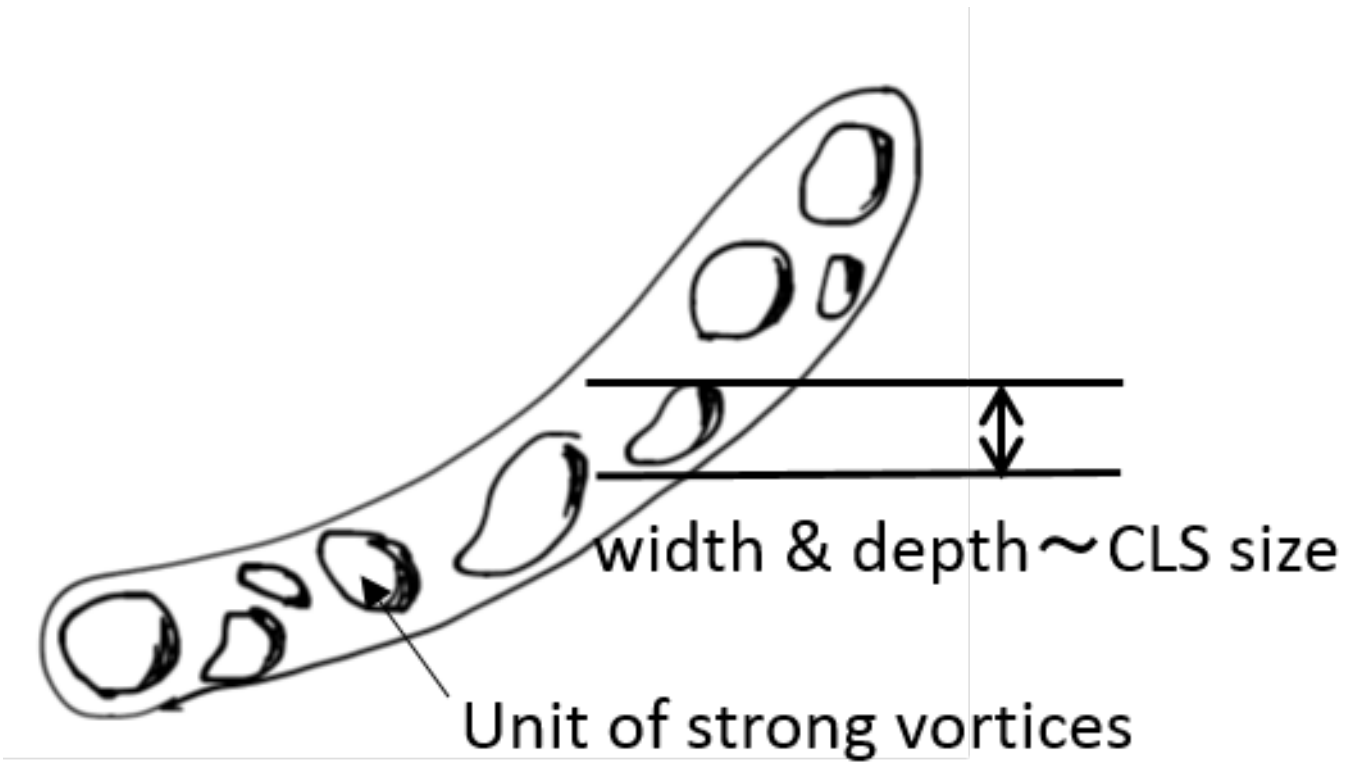

Fig. 1 Illustrative sketch of large chain-like structure named the serpentinely organized vortices (SOV). A SOV structure contains a mass of coherently assembled elementary vortices. Sizes of these element vortices are comparable or smaller than the CLS. The SOV structure looks like densely distributed many peas (vortices) contained in a pod (chain structure).

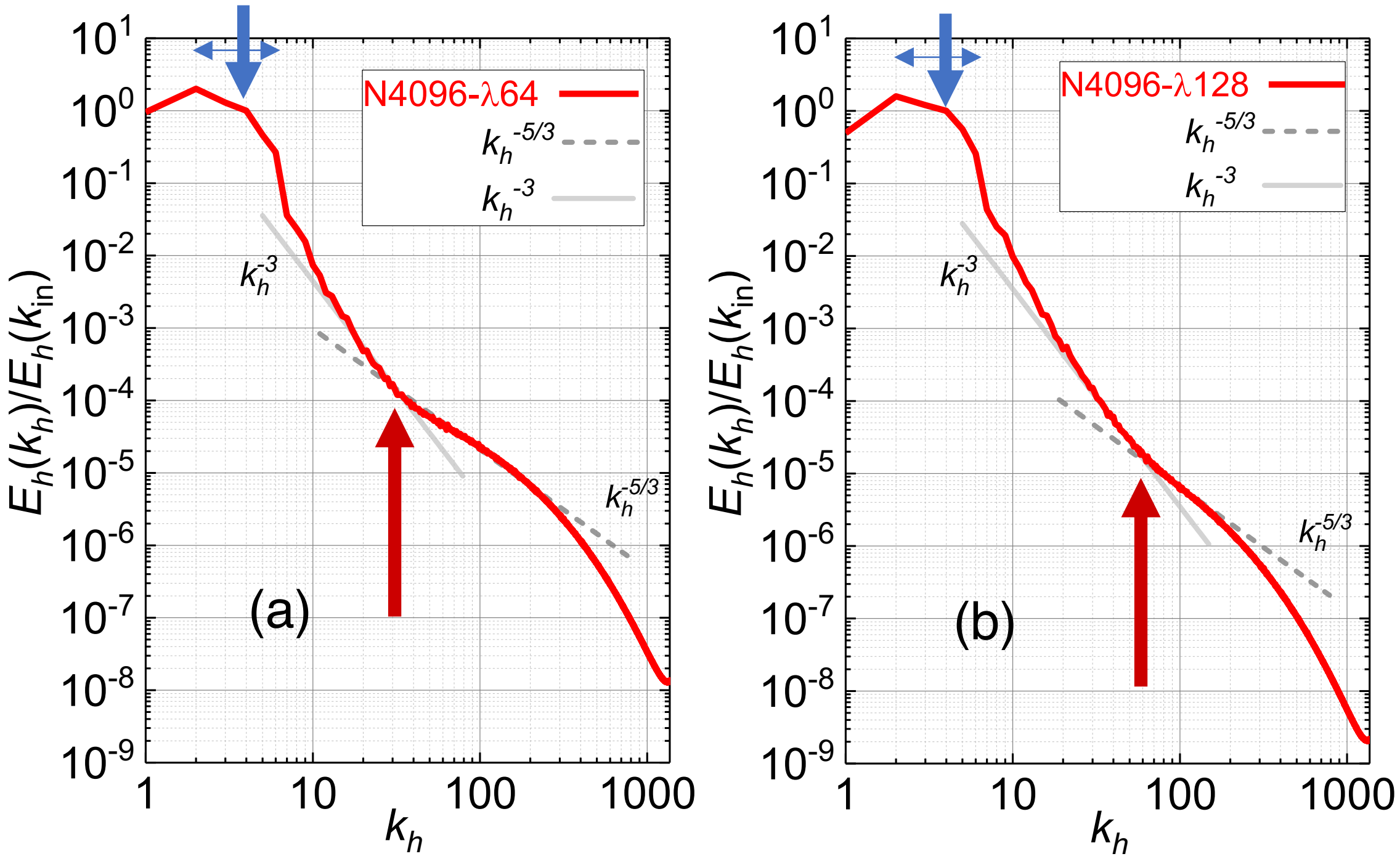

**Fig. 2** Energy spectra of horizontal velocity, $E_h(k_h) = 0.5\left(|\breve{u}_x(k_h)|^2 + |\breve{u}_y(k_h)|^2\right)$, normalized by $E_h$ at $k_h = k_{\mathrm{in}}$. Here, $\breve{u}_x(k_h) = \left(\sum_{k_z} \hat{u}_x^2(k_h, k_z)\right)^{1/2}$ for (a) N4096-λ64 and (b) N4096-λ128. The vertical (horizontal) blue arrows indicate the center (spread) of energy injection. The red arrows indicate half of the CLS wavenumber, i.e., $0.5k_{\mathrm{cr}}$.

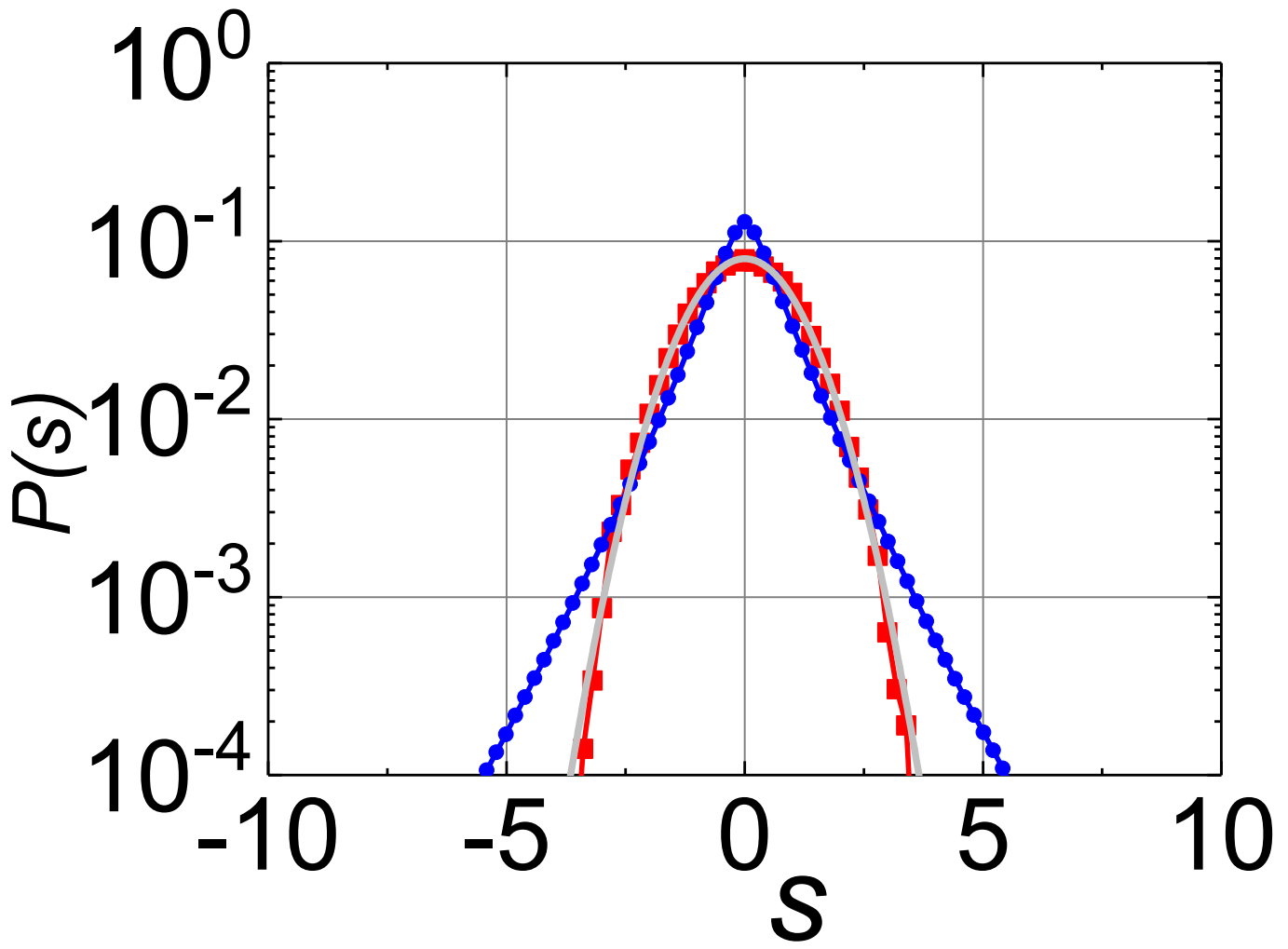

**Fig. 3** PDF of horizontal velocity $u_y$ (red curve with red squares) and $\Delta u_y(L_z/2,0)$ (blue curve with blue circles) for N4096-λ64, with the Gaussian distribution shown for reference (gray). Here and in the following figures, the abscissae of PDF and the color contours are normalized by each standard deviation ($\sigma$). $\Delta u_y(L_z/2,0)$ shows a wider tail, meaning stronger intermittency.

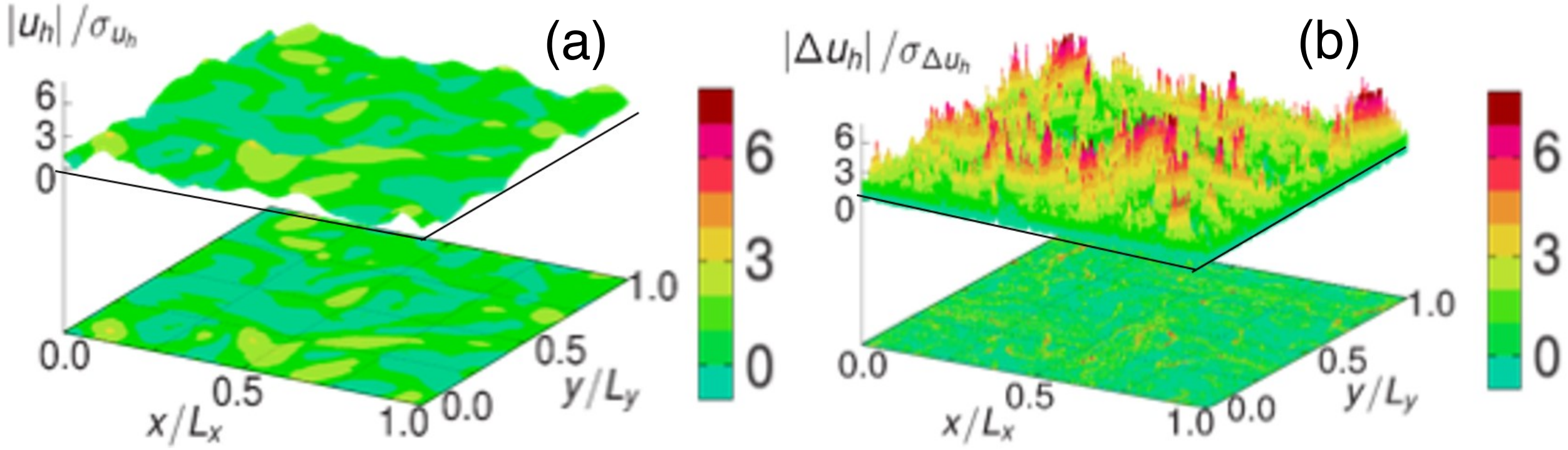

**Fig. 4** Typical in-plane contour snapshots of horizontal velocities and bird's-eye views for (a)$|\boldsymbol{u}_h|$ at $z = 0$ and (b) $|\Delta\boldsymbol{u}_h|$ for N4096-λ64. The color contours are normalized by the standard deviations of $u_y$, $\sigma(u_y)$ (or equivalently $\sigma(u_x)$) in (a) and $\sigma(\Delta u_y)$ (or $\sigma(\Delta u_x)$) in (b). The mean and standard deviation are 0.899 and 0.737, respectively, in (a) and 0.0374 and 0.0327, respectively, in (b). Note that $|\boldsymbol{u}_h| = \sqrt{u_x^2 + u_y^2}$ and $|\Delta\boldsymbol{u}_h| = \sqrt{\Delta u_x^2 + \Delta u_y^2}$. Large values appear in (b), corresponding to the wide tail in the PDF of $\Delta u_y(L_z/2,0)$.

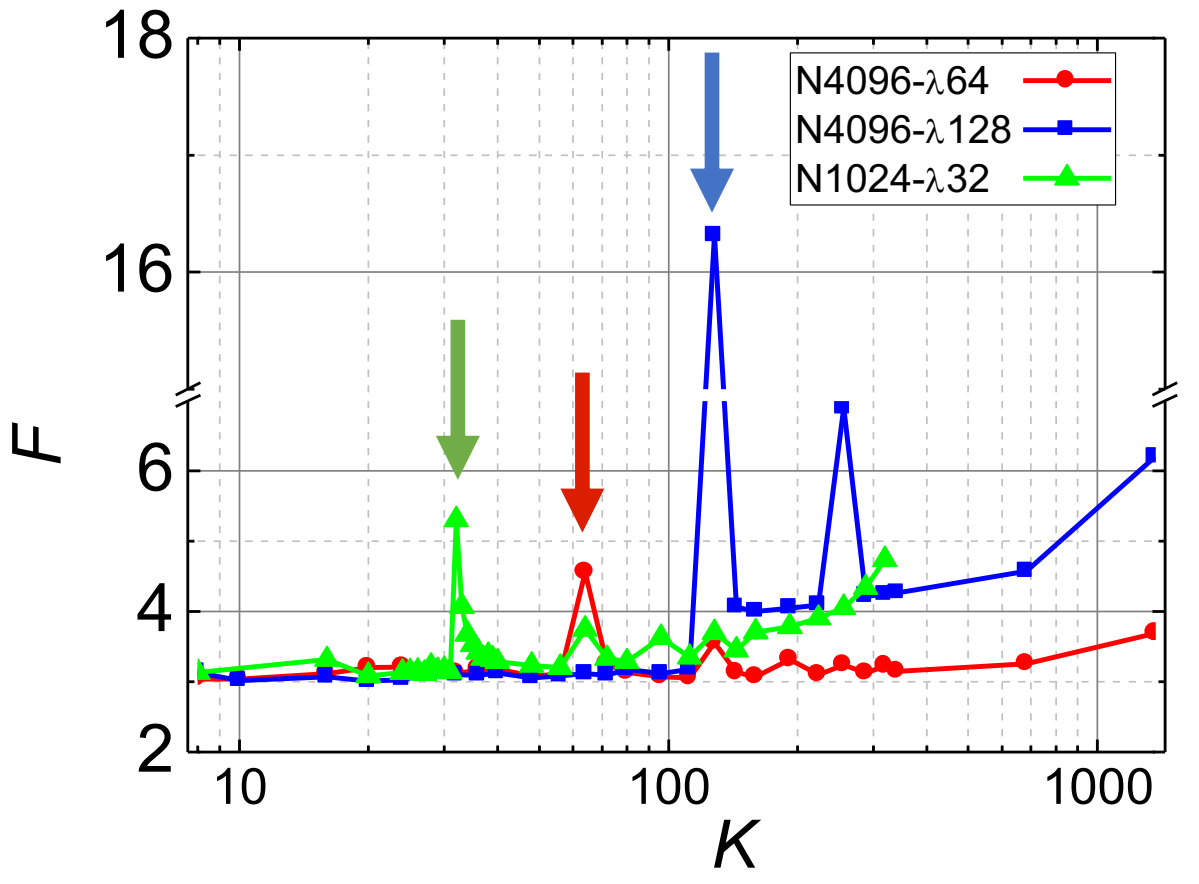

**Fig. 5** Flatness factors of the spherical-shell (band-pass)-filtered horizontal velocity $\tilde{u}_h^K$ with filter scale $K$ (see Appendix A1) for N4096-λ64, N4096-λ128, and N1024-λ32. Arrows indicate $k_{\mathrm{cr}}$ for each case (red, blue, and green, respectively). The prominent intermittency is always observed at the CLS.

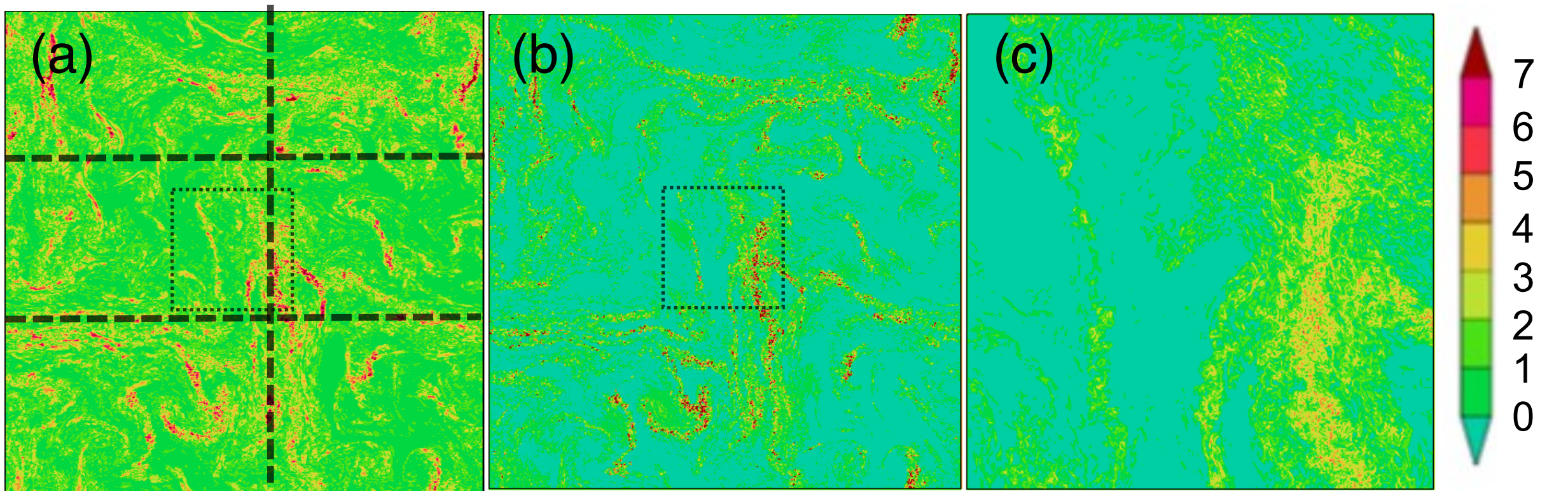

**Fig. 6** $xy$-plane cross section of (a) $|\Delta \boldsymbol{u_h}(L_z/2,0)|$ (see also Movie S1), (b) enstrophy $\zeta$, and (c) finite-time Lyapunov exponents $\Lambda$ at $z = 0$ for N4096-λ64. The region shown in (c) corresponds to the zoom-up of the dotted square areas in (a) and (b). The color contours are normalized by standard deviations. The mean and standard deviation are 2. 28 and 6.55, respectively, in (b) and 0.603 and 0.378, respectively, in (c). The distributions of extreme values indicated by red or dark yellow areas form SOV (chain structure) and the shapes are very similar among the three sub-figures.

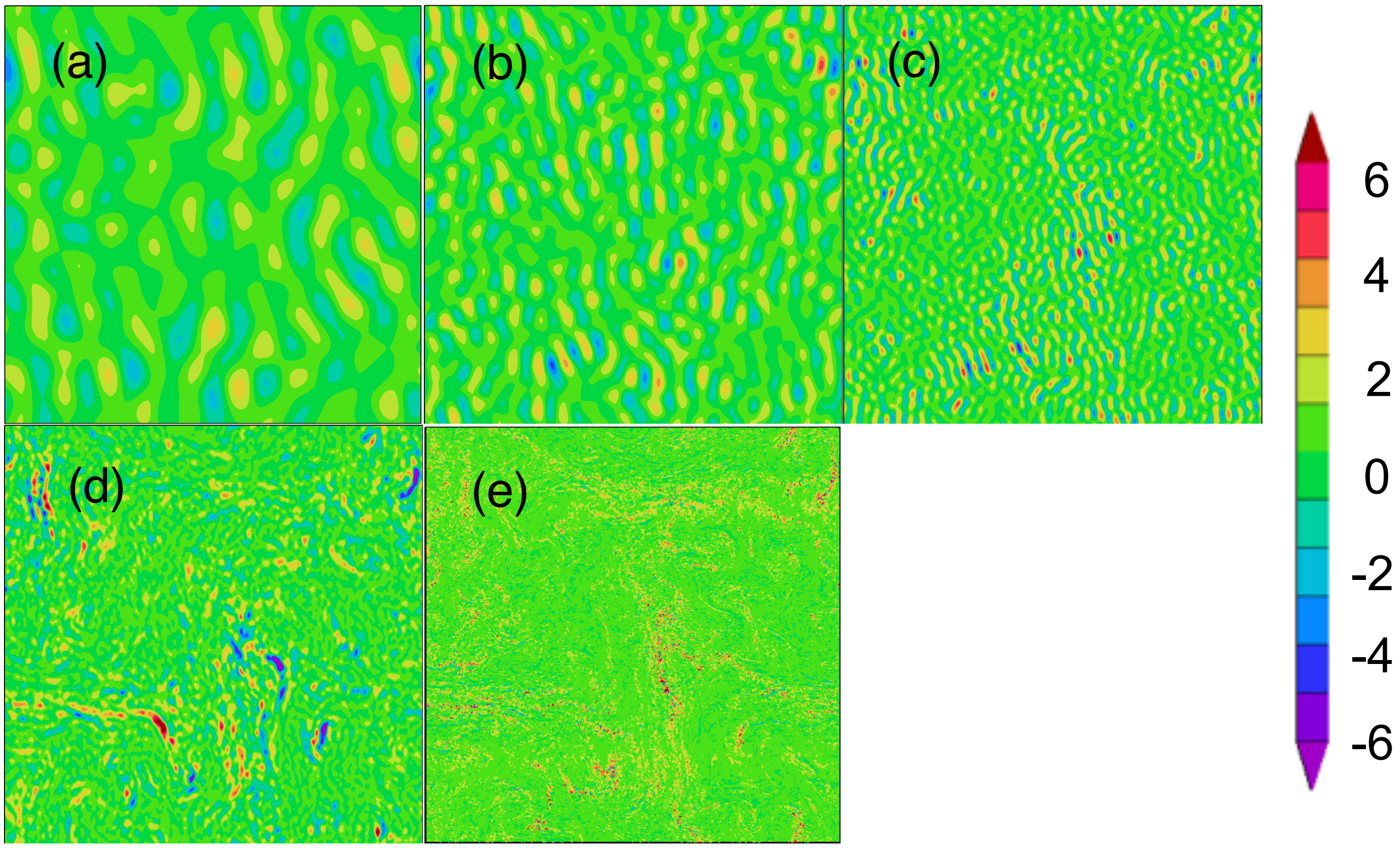

**Fig. 7** *xy*-plane cross section of low-pass filtered horizontally oriented vorticity $\omega_y$ (see Appendix A2) at $z = 0$ with threshold wavenumbers of (a) 10, (b) 20 (c) 40 (d) 80 and (e) 4096 for N4096-λ64. The color contours are normalized by the standard deviations. The characteristic large-scale SOV feature is observed only when the cross-over wavenumber is included, i.e., only in (d) and (e).

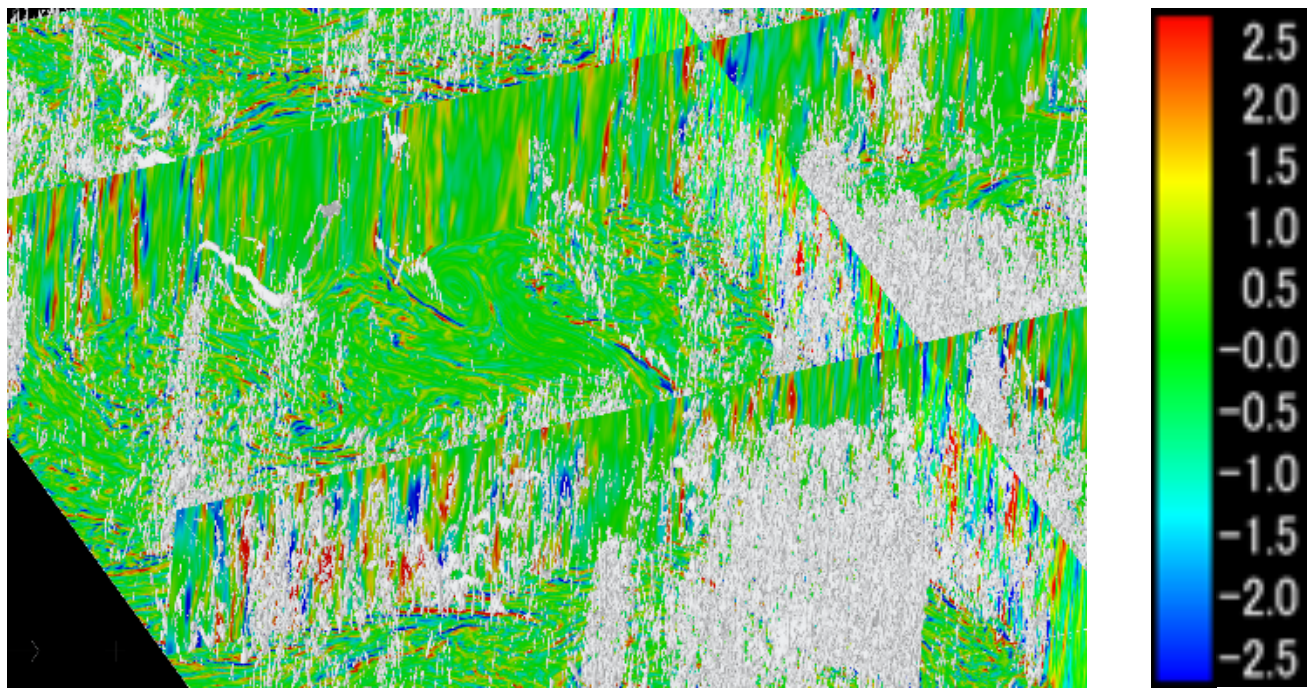

**Fig. 8** Distributions of the horizontally oriented vorticity (color contour) at three vertical walls whose top view coordinates are indicated by black dashed lines in Fig. 6(a) together with the color contour at the bottom plane at $z$=0 for N4096-λ64. The enstrophy is also plotted by white 3D iso-surface. The threshold of the white iso-surface was set at $m + 3\sigma$, where $m$ and $\sigma$ are the mean and standard deviation, respectively. The vertical real-space scale is doubled. For vorticity, calm regions are colored green in the planes, and active regions are colored red or blue (as in color scale bar). Large structures (SOV) consist of a mass of assembled tiny active spots and specks with sizes comparable to or smaller than the CLS. The bottom plane cross section corresponds to the cross section illustrated in Fig. 6(b) and the active SOV structure has one-to-one correspondence. See also Movie S2 for a temporal evolution of the structures.

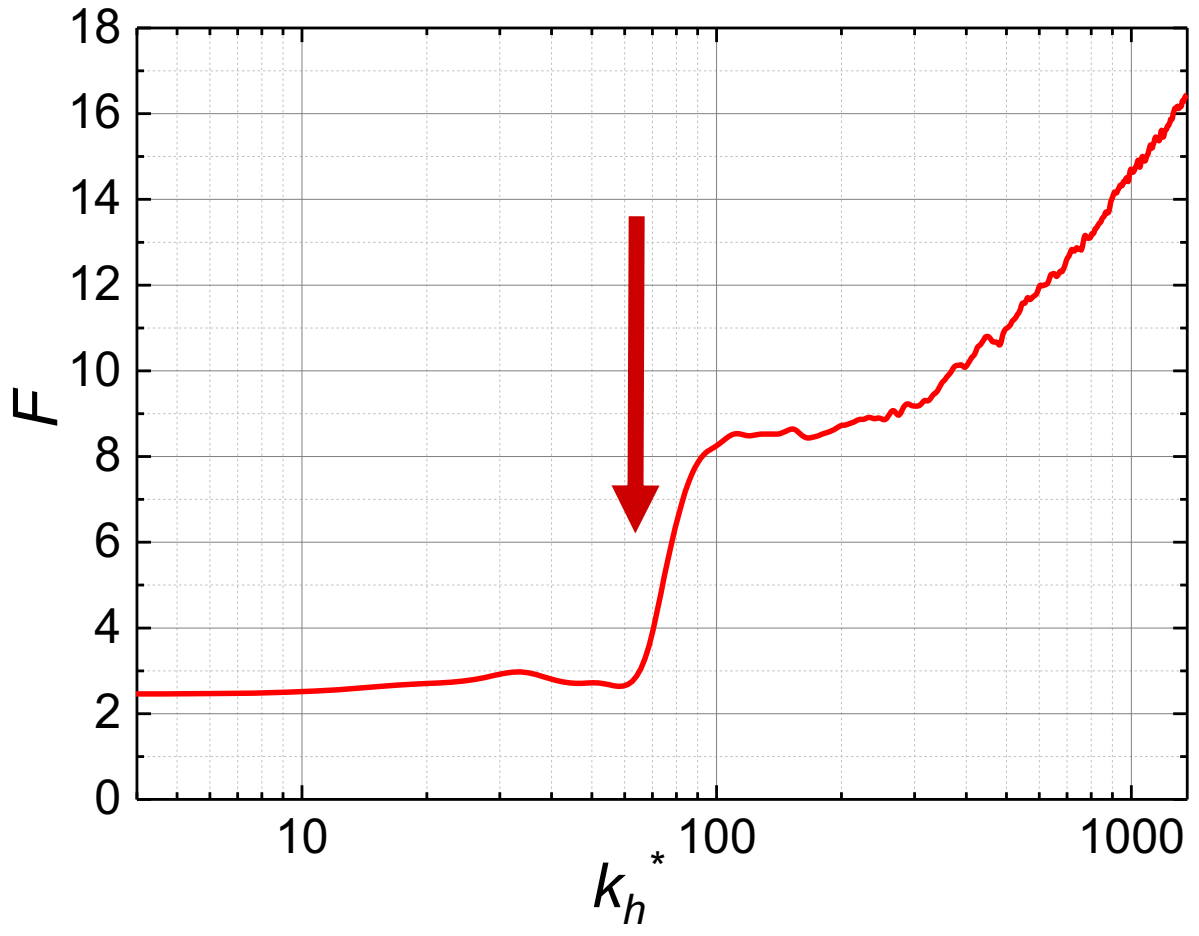

**Fig. 9** Flatness of the time series of $u_y$ estimated after frequency-resolved bandpass filtering at frequency ω (see Appendix A4) for N4096-λ64. The abscissa shows $k_h^* (\equiv \omega/u_{\mathrm{rms}}$, where $u_{\mathrm{rms}}$ is the root mean square of velocity in the whole simulation). The red arrow indicates the CLS wavenumber, $k_{\mathrm{cr}}$.

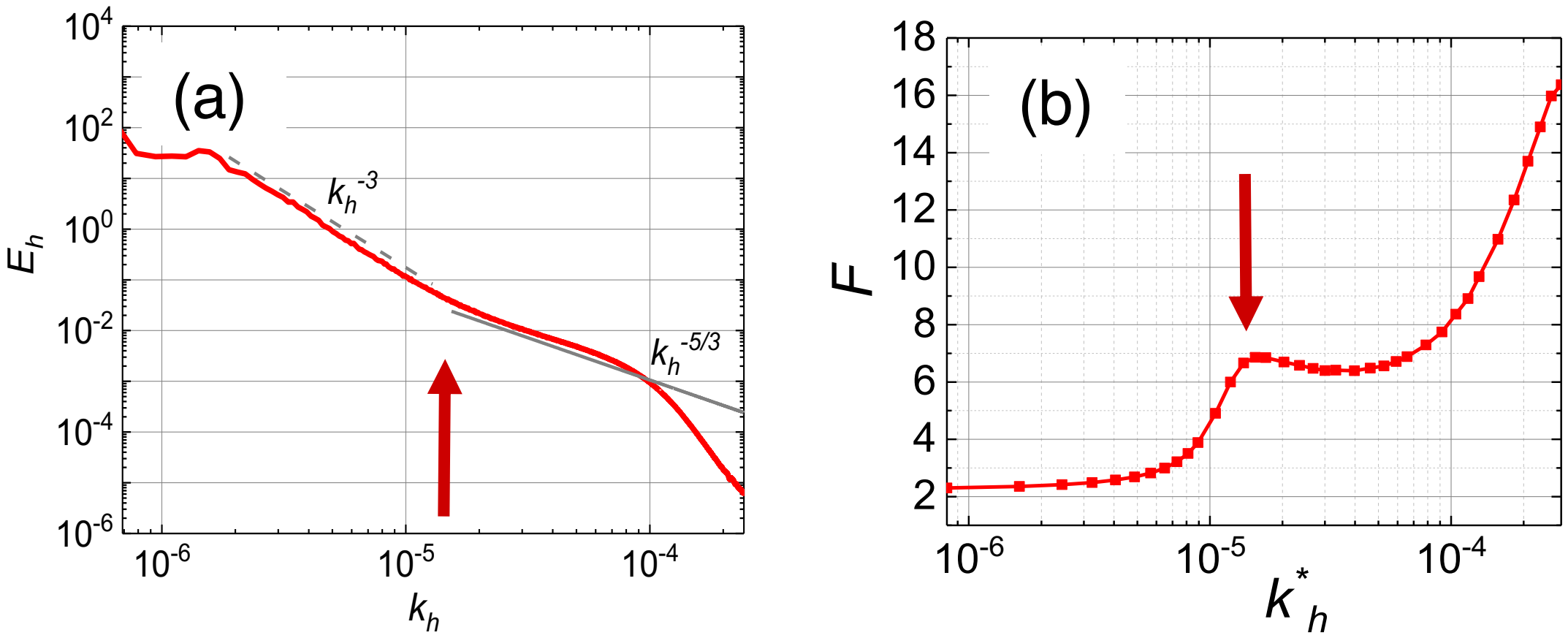

**Fig. 10** Results from global atmospheric simulation. (a) Energy spectra of horizontal wind at 200 hPa. (b) Flatness of horizontal velocities at 200 hPa estimated by frequency-resolved bandpass filter. The graphs were obtained by averaging the data at 1,024 measurement points, which were dispersed uniformly on the whole globe. Here, $k_h^*$ $(= 2\pi(Ut)^{-1}$, with $U$ chosen to be the root mean square of the horizontal velocity at 200 hPa for the entire global domain) is the effective wavenumber.

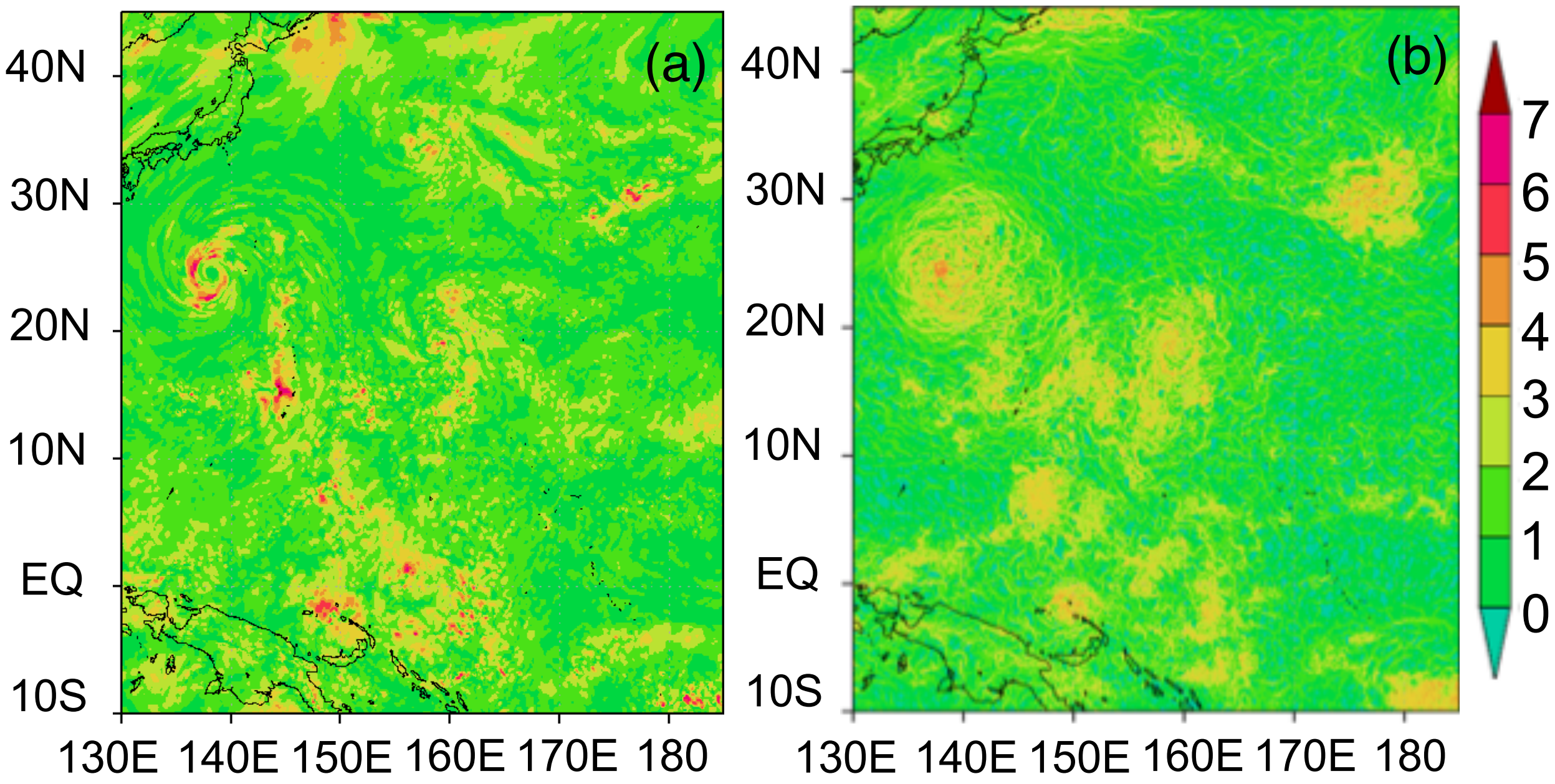

**Fig. 11** (a) Absolute value of the difference between the horizontal winds at 700 and 850 hPa $|\Delta \boldsymbol{u}_h| = |\boldsymbol{u}_h(700\text{hPa}) - \boldsymbol{u}_h(850\text{hPa})|$ (see also Movie S3). (b) Finite-time Lyapunov exponent $\Lambda$ at 700 hPa. The color contours are normalized by $\sigma(\Delta u_h)$ in (a) and $\sigma(\Lambda)$ in (b). The mean and standard deviation are 4.27 and 3.07, respectively, in (a) and $2.42\times10^{-5}$ $s^{-1}$ and $2.15\times10^{-5}$ $s^{-1}$, respectively, in (b). Active regions form similar structures in (a) and (b).

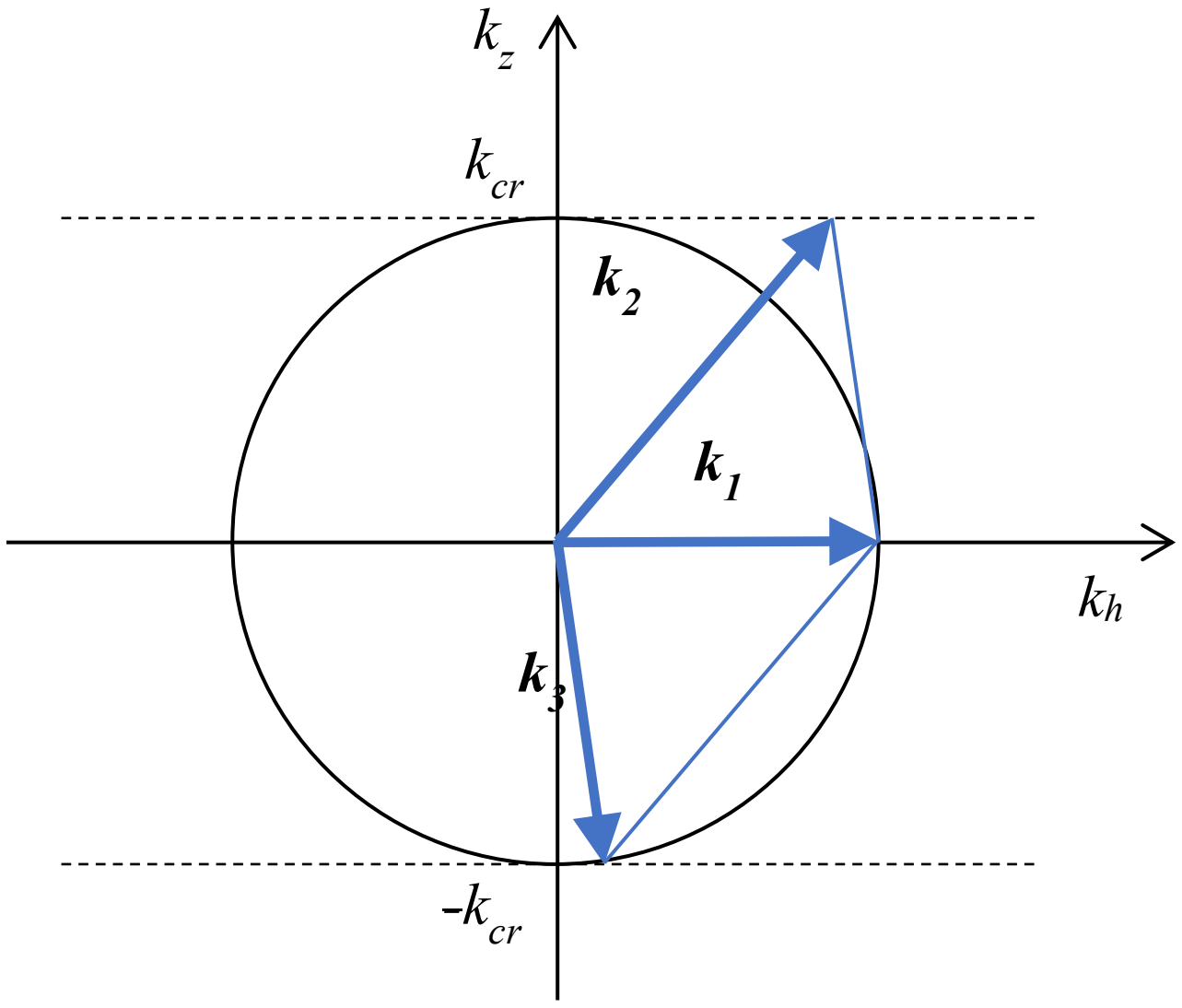

**Fig. 12** Schematic illustration of the formation mechanism of the colossal structure. Vector $\boldsymbol{k}_3$, whose horizontal wavenumber is small, is generated via mode coupling that satisfies momentum conservation.

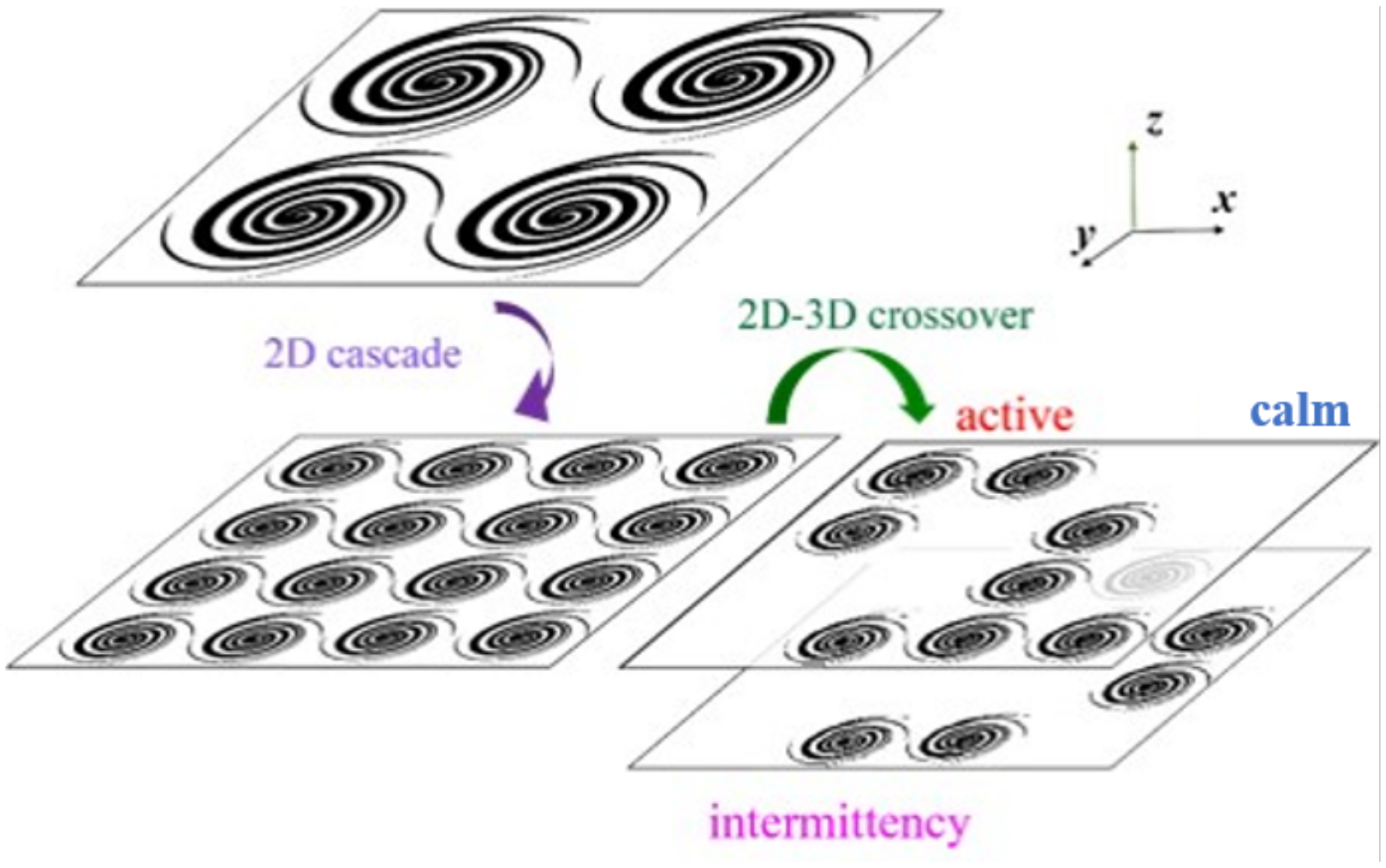

**Fig. 13** Schematic illustration of the emergence of the intermittency by phase space expansion at the 2D-3D crossover. In the 2D cascade, the self-similarity is approximately satisfied and the broken-up smaller vortices fill the space again. However, at the 2D-3D crossover, the active area is unable to fill the whole phase space anymore because of the expansion of the phase space in the $z$ direction. This generates active and calm regions with intermittency.

# APPENDIX A

**Finite-time Lyapunov exponent**

The Lyapunov exponent of a dynamical system is a measure that characterizes the rate of separation of infinitesimally close orbits in phase space. The inverse of the exponent can be a measure of the predictable time duration of the system. That is, the Lyapunov exponent is also a measure of unpredictability. Let us consider a particle at position $\boldsymbol{x}_0$ at time $t_0$ advected to position $\boldsymbol{x}_1$ at time $t_1$. The mapping operator $\boldsymbol{F}$ is defined as $\boldsymbol{F}_{t_0}^{t_1}(\boldsymbol{x}_0) = \boldsymbol{x}_1(t_1, t_0, \boldsymbol{x}_0)$. Two neighboring particles initially located around $\boldsymbol{x}_0$ at $t_0$ would be advected to $\boldsymbol{x}_1$ at $t_1$. The separation vector between the two particles would be stretched. The stretching matrix for initially infinitesimal separation can be written as:

$$\nabla\boldsymbol{F}_{t_0}^{t_1}(\boldsymbol{x}_0) \equiv \begin{pmatrix} \frac{\partial x_1}{\partial x_0} & \frac{\partial x_1}{\partial y_0} & \frac{\partial x_1}{\partial z_0} \\ \frac{\partial y_1}{\partial x_0} & \frac{\partial y_1}{\partial y_0} & \frac{\partial y_1}{\partial z_0} \\ \frac{\partial z_1}{\partial x_0} & \frac{\partial z_1}{\partial y_0} & \frac{\partial z_1}{\partial z_0} \end{pmatrix}. \tag{A1}$$

The finite-time Lyapunov exponent $\Lambda$ is then defined as [28]:

$$\Lambda = \frac{1}{|t_1 - t_0|} \ln\sqrt{\lambda_{\max}(\Delta)}, \tag{A2}$$

where $\lambda_{\max}(\Delta)$ is the maximum eigenvalue of the symmetric mapping matrix $\Delta$, which is defined as $\Delta \equiv \left[\nabla\boldsymbol{F}_{t_0}^{t_1}(\boldsymbol{x}_0)\right]^* \left[\nabla\boldsymbol{F}_{t_0}^{t_1}(\boldsymbol{x}_0)\right]$. The finite time $(t_1 - t_0)$ was chosen to be one-tenth of the turnover time of the largest possible eddy $T_e$, defined by $T_e = L_x/u_{\rm rms}$ for Fig. 6(c), where $L_x$ $(= 2\pi)$ is the horizontal size of the system, and $u_{\rm rms}$ is the root mean square of the horizontal velocity of the total system. For Fig. 11(b) and Movie S3, $T_e = 5$ days, which is a typical synoptic timescale and the integration time duration for the typhoon simulation. The exponents do not appreciably depend on the choice of $T_e$.

# APPENDIX B

**Three-dimensional homogeneous isotropic turbulence**

For comparison, the data from the DNS of 3D homogeneous isotropic turbulence (HIT) simulated with 512×512×512 grids [30] were analyzed. The Taylor microscale-based Reynolds number was 210.

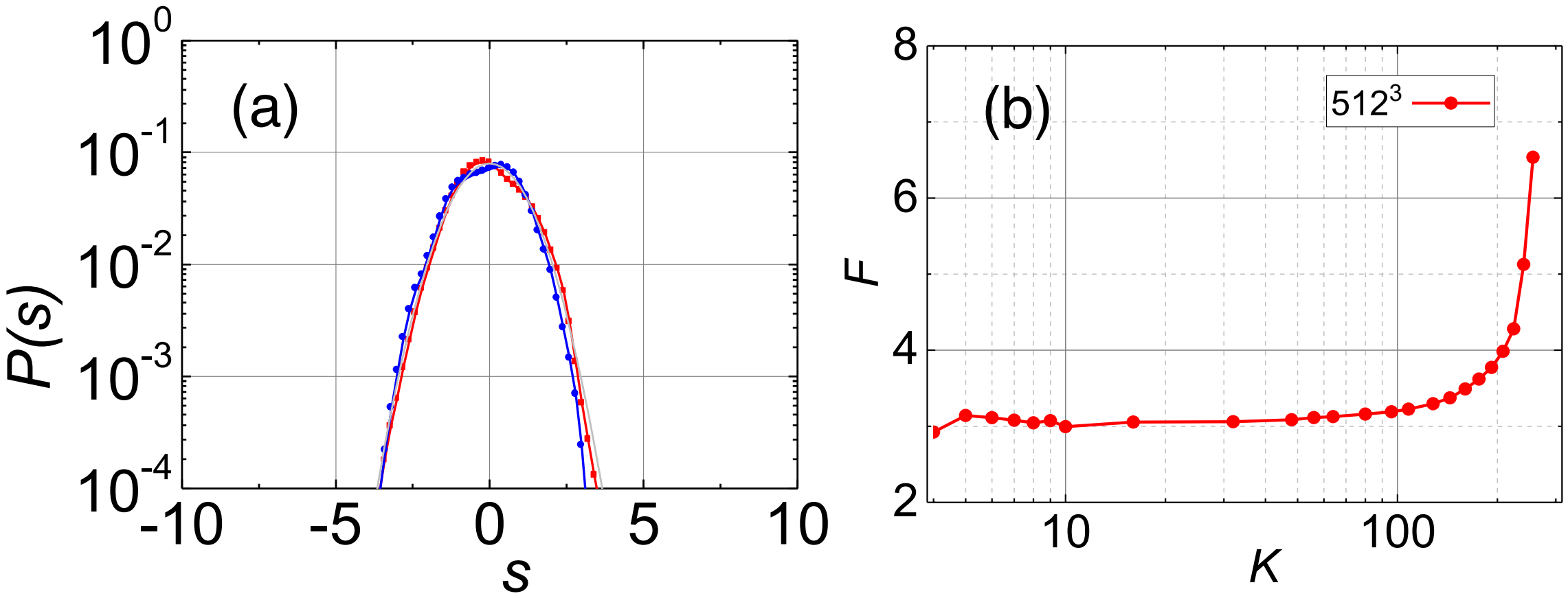

**Fig. B1** Properties of isotropic homogeneous 3D turbulence. (a) PDFs of horizontal velocity $u_y$ (red curve with red squares) and $\Delta u_y(L_z/2,0)$ (blue curve with blue circles). (b) Flatness factors of the spherical-shell (band-pass)-filtered horizontal velocity $\tilde{u}_h^K$ with filter scale $K$. Unlike Fig. 5, which shows the flatness factors in F3DF, strong intermittency is limited only in the dissipation scales. It should be noted that the Taylor microscale corresponds to $K = 27$ in (b).

# APPENDIX C

The present study uses various filters to extract the wavenumber-resolved fluid velocity to capture the role of the CLS throughout the entire energy cascade process.

**C1. Spherical-shell (band-pass) filter with window function for Fig. 5**

A window function depending on wavelength $|\boldsymbol{k}|$ with width $\Delta k$ is used for the spherical-shell filter:

$$\tilde{\boldsymbol{u}}^K(\boldsymbol{x},t) = \sum_{\boldsymbol{k}} \hat{G}^K(\boldsymbol{k})\hat{\boldsymbol{u}}(\boldsymbol{k},t)\exp(i\boldsymbol{k}\cdot\boldsymbol{x}), \tag{C1}$$

where

$$\hat{G}^K(\boldsymbol{k}) = \begin{cases} 1 & K-\Delta k/2 \le |\boldsymbol{k}| < K+\Delta k/2 \\ 0 & \text{otherwise} \end{cases}, \tag{C2}$$

where $\Delta k\ (= 2\pi/L_x = 1)$ is the smallest wavenumber in the discrete mesh for the simulations.

**C2. Low-pass filter with step function for Fig. 7**

In the low-pass filter, only modes with a wavenumber of less than *K* are retained:

$$\tilde{\boldsymbol{u}}^{<K}(\boldsymbol{x},t) = \sum_{\boldsymbol{k}} \hat{G}^{<K}(\boldsymbol{k})\hat{\boldsymbol{u}}(\boldsymbol{k},t)\exp(i\boldsymbol{k}\cdot\boldsymbol{x}), \tag{C3}$$

where

$$\hat{G}^{<K}(\boldsymbol{k}) = \begin{cases} 1 & |\boldsymbol{k}| < K \\ 0 & \text{otherwise} \end{cases}. \tag{C4}$$

**C3. Gauss-weighted filter for Fig. D1,**

The filtered velocity $\tilde{\boldsymbol{u}}^g(k_v)$ is defined by:

$$\tilde{\boldsymbol{u}}^g(\alpha,k_v)(\boldsymbol{x},t) = \sum_{\boldsymbol{k}} \hat{G}_{k_v}{}^g(\alpha,\boldsymbol{k})\hat{\boldsymbol{u}}(\boldsymbol{k},t)\exp(i\boldsymbol{k}\cdot\boldsymbol{x}), \tag{C5}$$

where

$$\hat{G}_{k_v}{}^g(\alpha,\boldsymbol{k}) = \begin{cases} \exp[-\{\alpha k_h\}^2] & |k_z| = k_v \\ 0 & \text{otherwise} \end{cases}, \tag{C6}$$

$k_h \left(= \left(k_x^2+k_y^2\right)^{1/2}\right)$ is the horizontal wavenumber, and $\boldsymbol{k} = (\boldsymbol{k}_h, k_z)$. This filter acts as a low-pass filter that extracts the modes with $k_h < 1/\alpha$ in wavenumber space and those with $\left((x^2+y^2)^{1/2} =\right) l_h < 1/\alpha$ in real space. The advantage of the Gauss-weighted filter is that the filter in the wavenumber space is equivalent to the filter in the real space, so that $\alpha$ picks up the characteristics of the length scale $1/\alpha$ in real space.

**C4. Frequency-resolved band-pass filter for Fig. 9 and Fig. 10(b)**

The band-pass filter in frequency space is defined as:

$$\tilde{\boldsymbol{u}}(\omega,\boldsymbol{x_0},t) = \sum_{\omega'} \hat{G}^{\omega}(\omega')\hat{\boldsymbol{u}}(\boldsymbol{x_0},\omega')\, e^{i\omega' t}, \tag{C7}$$

where

$$\hat{G}^{\omega}(\omega') = \exp\left(-\left(\frac{\omega'-\omega}{\Delta\omega}\right)^2\right), \tag{C8}$$

where $\Delta\omega$ is set to $2\pi/(\alpha T_a)$, where $T_a$ is the time length for analysis (i.e, sampling time) and the factor $\alpha$ is set to 0.125. $T_a$ is 16.4 for the DNS of F3DF and $4.32\times10^5$ s (5 days) for the atmospheric simulation. Flatness is obtained as:

$$F(\omega, \boldsymbol{x_0}) = \frac{\frac{1}{N_T}\sum_t \{\tilde{u}_x(\omega, \boldsymbol{x_0}, t)\}^4}{\left\{\frac{1}{N_T}\sum_t \{\tilde{u}_x(\omega, \boldsymbol{x_0}, t)\}^2\right\}^2}. \tag{C9}$$

Averaging over $\boldsymbol{x}_0$ to reduce statistical error yields:

$$F(\omega) = \frac{1}{N_{x_0}}\sum_{\boldsymbol{x_0}} F\,(\omega, \boldsymbol{x_0}). \tag{C10}$$

The effective wavenumber $k^*$ is introduced to transform $\omega$ by $k^* = \omega/U$. Here, the representative velocity $U$ is given by the average advection velocity, defined as:

$$U = \frac{1}{N_{x_0}}\sum_{\boldsymbol{x_0}} U(\boldsymbol{x_0}), \tag{C11}$$

where

$$U(\boldsymbol{x_0}) = \sqrt{\frac{1}{N_T}\sum_t \{u_x^2(\boldsymbol{x}_0, t) + u_y^2(\boldsymbol{x}_0, t)\}^2}\ . \tag{C12}$$

# APPENDIX D

**Flatness of Gaussian-filtered velocity**

Here, the flatness of velocity $\tilde{\boldsymbol{u}}^g(k_v)$ obtained by the Gaussian filter with width $1/\alpha$ centered at $k = 0$ in momentum space is shown.

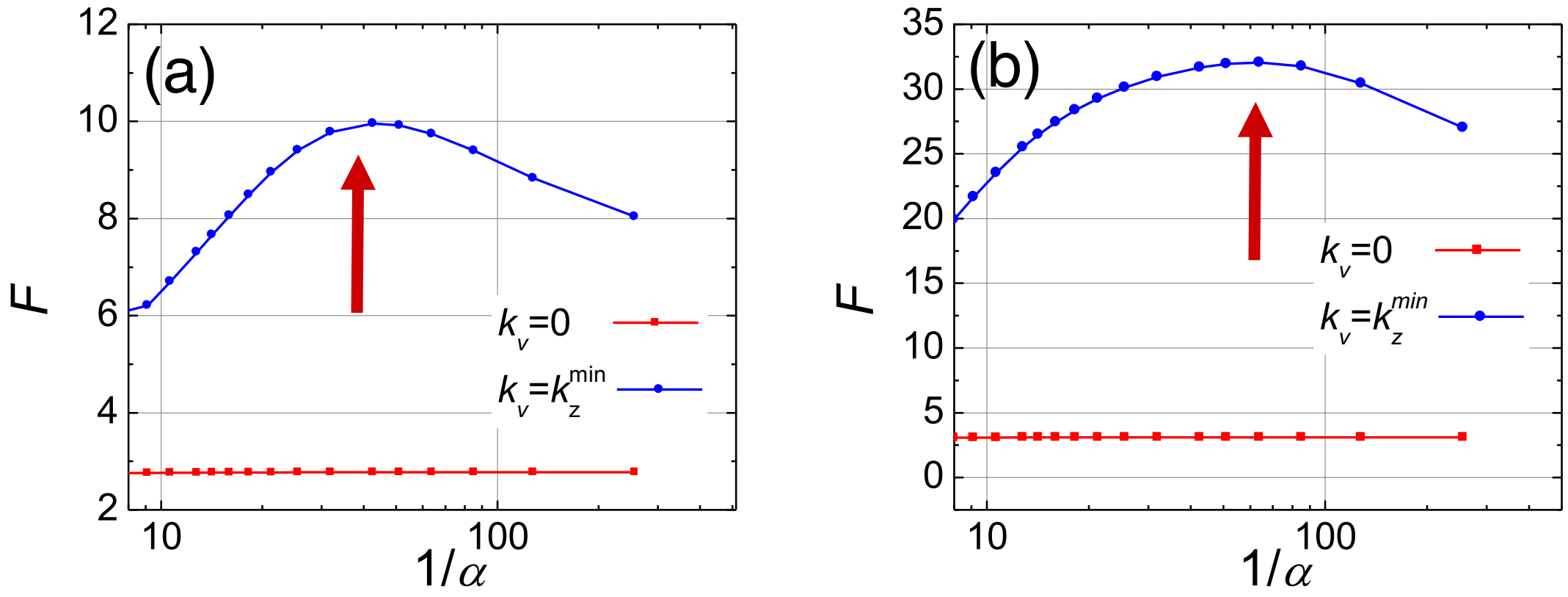

**Fig. D1** Flatness of Gaussian-filtered velocity $\tilde{\boldsymbol{u}}^g(k_v)$ for F3DF with vertical wavenumber $k_v = 0$ and $k_v = k_z^{min}$ for (a) N4096-λ64 and (b) N4096-λ128. The intermittency is pronounced only for $k_v = k_z^{min}$ around $1/\alpha = 0.5k_{\rm cr}$, indicating that it originated from the dimensional crossover.

# APPENDIX E

**Role of mode coupling in large SOV structure generation with intermittency**

Velocities band-pass-filtered at $k_{\mathrm{cr}}$ exhibit strong intermittency, as shown in Fig. 5(a). There are multiple modes at $|\boldsymbol{k}| = k_{cr}$, e.g., $k = (0,0,\pm k_{\mathrm{cr}})$ and $(\pm k_{\mathrm{cr}}, 0,0)$. Figure E1 shows the role of mode coupling for the intermittency at $|\boldsymbol{k}| = k_{cr}$. The PDFs of the velocities that have a single mode with $|\boldsymbol{k}| = k_{cr}$ exhibit Gaussian-like distributions (Figs. E1(a) and E1(b)), while those containing contributions from multiple modes clearly exhibit non-Gaussian distributions (Fig. E1(c)). However, if different modes follow independent Gaussian distributions, their linear combinations also follow a Gaussian distribution. Therefore, the non-Gaussianity supports the existence of correlations between different modes. Mode coupling is essential for the strong intermittency at the CLS, as discussed in the main text.

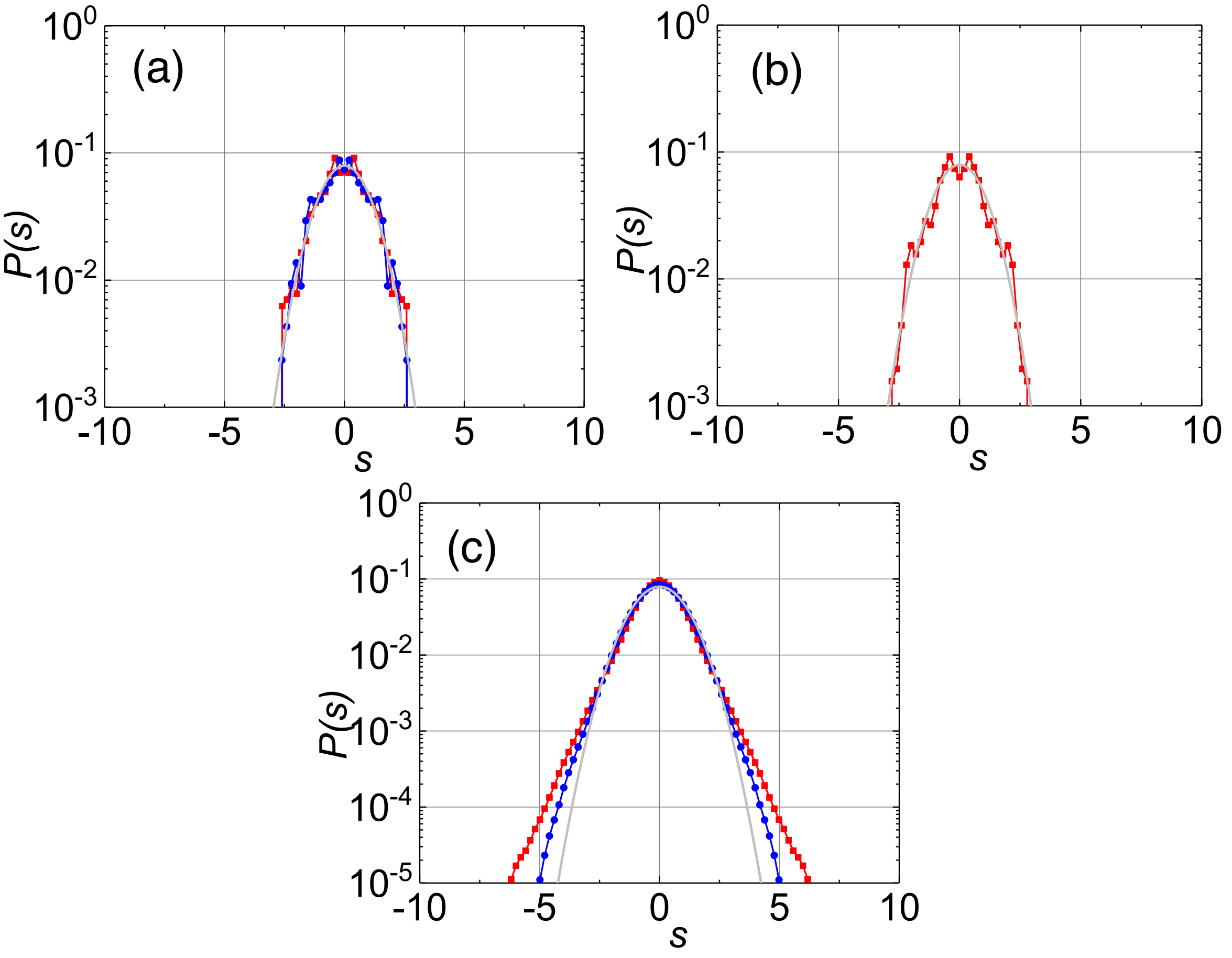

**Fig. E1** PDFs of mode-selected velocities. Horizontal (red)/vertical (blue) mode-extracted velocities $u_{y|z}(\boldsymbol{k};\boldsymbol{x},t) = \sum_{k'=k} \hat{u}_{y|z}(\boldsymbol{k}',t)\exp(i\boldsymbol{k}'\cdot\boldsymbol{x})$ consist of several modes (defined below). These PDFs are compared with the Gaussian distribution (grey). (a) $\boldsymbol{k} = (\pm k_{\mathrm{cr}}, 0,0)$, (b) $\boldsymbol{k} = (0,0,\pm k_{\mathrm{cr}})$, and (c) $\boldsymbol{k} = \left(k_x, k_y, \pm k_{\mathrm{cr}}\right)$ with $k_{\mathrm{cr}} - 1/2 \le \left(k_x^2 + k_y^2 + k_z^2\right) < k_{\mathrm{cr}} + 1/2$. $u_z(\boldsymbol{k};\boldsymbol{x},t) = 0$ is satisfied due to the divergence-free condition in (b). Only (c) shows clear deviations from the Gaussian distribution, i.e., clear intermittency.

**SUPPLEMENTARY MATERIALS**

**Movie S1**
Dynamics of intermittency and colossal structure in a 2D cross section for velocity increment for $xy$-plane cross section of $|\Delta \boldsymbol{u_h}(L_z/2{,}0)|$.

**Movie S2**
3D view of dynamics of intermittency and colossal structure in direct numerical simulation of flattened 3D fluid: distribution of enstrophy (white) and horizontally oriented vorticity (color contour). The vertical scale is elongated two-fold. The animation is for the region in Fig. 8.

**Movie S3**
Intermittency in global atmospheric simulation: absolute value of the difference between horizontal winds at 700 and 850 hPa.